\documentclass{article}

\usepackage[preprint]{css}

\usepackage[utf8]{inputenc} 
\usepackage[T1]{fontenc}    
\usepackage{hyperref}       
\usepackage{url}            
\usepackage{booktabs}       
\usepackage{amsfonts}       
\usepackage{nicefrac}       
\usepackage{microtype}      
\usepackage{amsmath, amssymb, graphicx}
\usepackage{algorithm}
\usepackage{algorithmic}
\usepackage{algorithmic}
\title{Symmetric Policy Design for Multi-Agent Dispatch Coordination in Supply Chains}

\author{%
  Sagar~Sudhakara \\
  Los Angeles, CA 90007 \\
  \texttt{sagarsudhakara@gmail.com} \\
}

\begin{document}

\maketitle





\begin{abstract}
We study a decentralized dispatch coordination problem in a multi-agent supply chain setting with shared logistics capacity. We propose symmetric (identical) dispatch strategies for all agents, enabling efficient coordination without centralized control. Using a common information approach, we derive a dynamic programming solution that computes optimal symmetric dispatch strategies by transforming the multi-agent problem into a tractable dynamic program on the agents' common information state. Simulation results demonstrate that our method significantly reduces coordination cost compared to baseline heuristics, including belief-based strategies and an always-dispatch policy. These findings highlight the benefits of combining symmetric strategy design with a common information--based dynamic programming framework for improving multi-agent coordination performance.
\end{abstract}

\section{Introduction}

Coordinating distributed supply chain operations is increasingly crucial as logistics resources are shared. A common challenge arises when multiple warehouses must schedule shipments through a limited resource, such as a fixed number of dispatch slots. Without coordination, two extremes may occur: (a) \textit{Over-subscription}, where multiple facilities dispatch at the same time, exceeding capacity and causing queues; and (b) \textit{Under-utilization}, where all facilities wait, leading to wasted capacity and delayed deliveries. Poor coordination in dispatch schedules often causes truck congestion and delays [2].
Traditional solutions often rely on centralized scheduling or simple heuristics. However, centralized control may not scale, and sharing sensitive demand data can be impractical. Multi-agent systems offer a promising alternative for decentralized yet cooperative scheduling, though achieving near-optimal coordination remains challenging, especially under uncertainty about other agents' urgency or load [1,6].

This work formulates a multi-agent dispatch coordination problem where distributed agents share limited logistics capacity. Each agent independently decides whether to utilize a dispatch opportunity, aiming to minimize system inefficiency. We impose symmetry by requiring agents to follow identical decision rules based on their private urgency and shared observations. This ensures fairness, reduces complexity, and enables decentralized coordination.

We adopt the common information approach from decentralized stochastic control to solve this problem [3]. By leveraging shared history (e.g., past dispatches), agents learn about each other's states and coordinate actions. Our contributions include:
\begin{itemize}
    \item Formulating the dispatch scheduling problem with capacity constraints as a team decision problem using symmetric strategies, a novel application in supply chain logistics.
    \item Developing a dynamic programming framework with common information to derive the optimal symmetric dispatch policy, minimizing expected total cost.
    \item Presenting simulation results for small supply chains, showing significant reductions in collision and waiting costs compared to heuristic strategies like belief-based thresholds and always-dispatch.
\end{itemize}

We discuss the practical implications: the symmetric policy ensures fair access to shared resources, is scalable, and can be applied to larger networks. We also outline extensions to handle more complex scenarios.

The remainder of the paper is organized as follows: Section II defines the dispatch coordination problem; Section III details the common-information dynamic programming methodology; Section IV presents experimental results and analysis; Section V discusses the significance and practical considerations; and Section VI concludes with future research directions.

\section{Problem Formulation}
We consider a supply chain scenario with $N$ agents (e.g., warehouses) sharing a dispatch resource, such as a fleet of trucks or a fixed number of dispatch slots. At most one dispatch can be processed per time slot. Each agent $i \in \{1, \dots, N\}$ decides whether to dispatch ($u^i_t = 1$) or wait ($u^i_t = 0$) at each time step $t = 1, 2, \dots, T$. If exactly one agent dispatches, the shipment is sent successfully. If multiple agents dispatch, a collision occurs, and only one succeeds. If no agent dispatches, the slot goes idle.

We define the cost function $k_t$ for each time slot as follows:
\[
k_t = 
\begin{cases} 
0 & \text{if exactly one agent dispatches at time } t, \\
1 & \text{otherwise (idle slot or collision)}.
\end{cases}
\]
Our objective is to minimize the expected total cost over the horizon $T$, i.e., minimize 
\[
J = \mathbb{E}\left[\sum_{t=1}^T k_t\right],
\]
which is equivalent to maximizing the number of successful dispatches. Each agent has a private urgency state $M^i$ (e.g., High or Low urgency). The private state is known only to the agent and does not change over time, though agents’ beliefs about each other evolve. We assume that agents share a common history of past actions, denoted $C_t = \{u^1_{1:t-1}, \dots, u^N_{1:t-1}\}$, which is common knowledge.

At each time $t$, agent $i$ decides whether to dispatch based on its private urgency $M^i$ and the common history $C_t$. We assume symmetric strategies, meaning all agents use the same decision rule $g_t$ that maps $(C_t, M^i)$ to a probability $\delta^i_t$ of dispatching [7]:
\[
g_t(C_t, M^i) \in [0, 1].
\]
The agents randomize their dispatch decisions according to this probability, ensuring that symmetry holds: for two agents with the same private state and observing the same common history, the dispatch probabilities are identical.

Our aim is to find the optimal symmetric strategy $g^*$ that minimizes the expected total cost:
\[
J(g^*, g^*, \dots, g^*) \le J(h, h, \dots, h) \quad \text{for any other symmetric policy } h.
\]
This results in a decentralized decision problem with imperfect information ~\cite{sudhakara2023symmetric}. Solving this problem directly is challenging due to the exponential growth of the joint state space and history. However, by leveraging the common information and symmetry, we can convert it into a tractable dynamic program, as outlined next.

\section{Methodology: Common Information Dynamic Programming}
To optimize the symmetric dispatch strategy, we use the common information approach [4,5], which views the problem from a central coordinator’s perspective. The coordinator observes only the common history $C_t$, not the private urgency states $M^i$ of the agents. The coordinator selects a prescription $\Gamma_t$, mapping private states to action probabilities, which all agents follow. For example, if an agent’s urgency is High, it dispatches with probability $p$, and if Low, it dispatches with probability $q$. The symmetry ensures that all agents follow the same policy.

The coordinator’s choice of $\Gamma_t$ depends on the common history $C_t$, and we define a coordination strategy $d = \{d_1, \dots, d_T\}$, where $d_t(C_t)$ selects the prescription $\Gamma_t$ at time $t$. This strategy is directly related to the agents’ strategy $g_t = d_t(C_t)$.

We define the belief at time $t$ as $\Pi_t$, which represents the probability distribution over agents' private states. For instance, in the $N=2$ case with binary urgency, $\Pi_t$ can be summarized by $\pi_t = P(\text{other agent is High} \mid C_t)$. The belief is updated using Bayes’ rule as actions are observed.

The dynamic program (DP) aims to find the optimal prescription $\Gamma_t$ at each time step that minimizes the expected total cost. The Bellman recursion for the value function $V_t(b)$, representing the optimal cost-to-go, is as follows:

\[
V_t(b) = \min_{\Gamma \in B} \left\{ \mathbb{E}_{U \sim \Gamma^N} \left[ k_t \mid \Pi_t = b, \Gamma \right] + \mathbb{E}_{U \sim \Gamma^N} \left[ V_{t+1}(b') \mid \Pi_t = b, \Gamma \right] \right\},
\]
where $b'$ is the updated belief after applying $\Gamma$ and observing $U$. The problem is solved backward from $t=T$ to $t=1$. At $t=T$, $V_{T+1} \equiv 0$.
The algorithm is as follows:

\begin{algorithm}[h!]
\caption{Dynamic Programming for Symmetric Dispatch Strategy}
\begin{algorithmic}[1]
\STATE Initialize belief $\Pi_1$
\FOR{t = T to 1}
    \STATE Compute the optimal policy $\Gamma_t$ by solving the Bellman equation
    \STATE Update belief: $\Pi_{t+1} = \mathcal{T}(\Pi_t, \Gamma_t, \text{observed } U_t)$
\ENDFOR
\STATE Output optimal coordination strategy $d^* = \{d_1^*, \dots, d_T^*\}$
\end{algorithmic}
\end{algorithm}

The DP finds a balance between collision risk and idle time, adjusting $\Gamma_t$ based on the current belief. For instance, if most agents are likely urgent, the prescription becomes conservative to avoid collisions. Conversely, if fewer agents are urgent, the prescription encourages dispatch to avoid idle time. This leads to a self-coordinating system.

\begin{figure}[h!]
\centering
\includegraphics[width=0.6\textwidth]{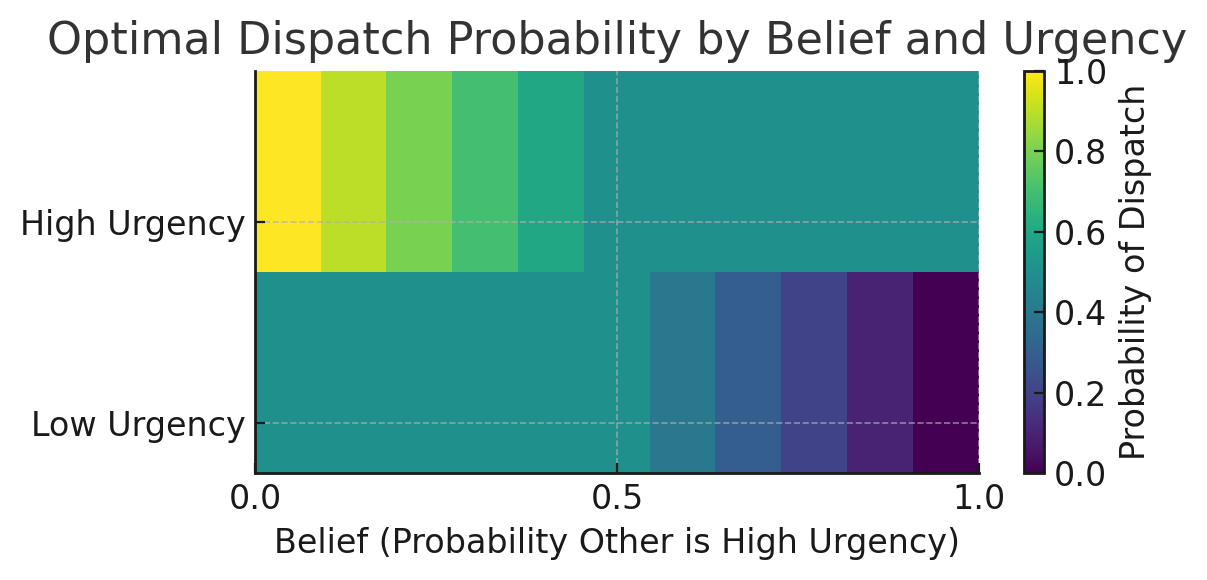}
\caption{Optimal symmetric dispatch policy probabilities as a function of agent urgency and the belief that other agents are high urgency. The color represents the probability of dispatching under the optimal policy (yellow = 1.0, dark blue = 0.0). High-urgency agents dispatch more often when they believe others are unlikely to dispatch (left side), but reduce probability as their belief in others' high urgency increases (right side). Low-urgency agents are less likely to dispatch if they believe another agent is certainly high urgency (right side), but will dispatch with moderate probability when they suspect others will not dispatch (left side).}
\end{figure}

The symmetric policy, visualized in Figure 1, shows that high-urgency agents dispatch with high probability when they believe others are not urgent, but reduce their dispatch probability to around 50\% when they believe others are also urgent. This tie-breaking leads to a 50\% chance of either a successful dispatch (cost 0) or a collision/idle slot (cost 1).

Low-urgency agents behave oppositely: if they believe another agent is likely to dispatch (high belief of other being urgent), they refrain from dispatching. If they believe no one else will dispatch (low belief), they dispatch with moderate probability (about 0.5) to avoid both agents idling or colliding. This strategy reflects a rational coordination policy, where urgent agents yield to each other to avoid clashes, and less-urgent agents dispatch just enough to keep the system running.

\section{Experiments}
We conducted numerical experiments on synthetic supply chain scenarios to illustrate the proposed symmetric strategy and compare it with baseline approaches. We primarily examine the case of $N=2$ agents (warehouses) in detail, since it allows easy visualization of beliefs and outcomes. Each simulation involves a time horizon $T$ (typically 10 slots for illustration) and random initial urgency states for each agent. High urgency (H) is assigned with some probability (load level), and Low urgency (L) otherwise. Agents do not directly know each other’s state, but they start with a prior belief (e.g., “50\% chance the other is High” if that’s the initial load probability). We implemented the common information dynamic program for $N=2$ with binary urgency by discretizing the prescription probabilities in increments of 0.05 (21 possible probability levels from 0 to 1). This yields a finite DP state-space of belief values (we also discretize the belief continuum $[0,1]$ into a grid of 5\% steps for tractability). The optimal symmetric strategy $g^*$ was obtained by backward induction on this discretized model. We then simulated the strategy forward under various initial conditions. For baselines, we consider:

\begin{itemize}
    \item \textbf{Belief-Threshold Heuristic:} a strategy where each agent uses simple threshold rules on the belief of others’ urgency, inspired by the heuristic policies in the IoT multi-access study ~\cite{sudhakara2023symmetric}. In our adaptation, an agent might behave as follows: “If I strongly believe another agent is urgent (above threshold $x$), I will wait (not dispatch). If I strongly believe no one else is urgent (below threshold $y$), I will dispatch. If in between, I dispatch with some fixed probability $z$.” By tuning $(x,y,z)$ one can obtain more aggressive or passive variants of this policy. We tested several such combinations (including extremes like always dispatch or always wait as special cases).
    
    \item \textbf{Always Dispatch (Aggressive):} a degenerate heuristic where every agent dispatches at every opportunity regardless of belief or urgency. This represents a failure of coordination (agents act purely selfishly or fear missing a slot).
    
    \item \textbf{Always Wait:} an extreme case where every agent always defers dispatch, resulting in all slots being idle. It was considered in analysis as an extreme case, though we do not plot it because it is obviously very poor unless one agent eventually breaks the tie.
\end{itemize}

To evaluate performance, we measure the cumulative cost over $T$ slots under each policy, averaged over many random trials (to account for the stochastic nature of strategies and initial uncertainties). Lower cost is better (0 would mean perfect coordination every slot; $T$ would mean failure every slot). We examine three representative initial load conditions:

\begin{itemize}
    \item \textbf{Light Load:} e.g., each agent has only 20\% chance to be High urgency. Likely 0 or 1 agents are truly in need at a time.
    
    \item \textbf{Moderate Load:} e.g., 50\% chance High for each (independent). Roughly one agent urgent on average, but sometimes two or none.
    
    \item \textbf{Heavy Load:} e.g., 80\% chance High for each. Often both agents are urgent simultaneously, competing for slots.
\end{itemize}

\textbf{Belief Evolution:} We illustrate how agents update their beliefs and coordinate using our symmetric strategy. Consider two agents, A and B, with true states A: High urgency, B: Low urgency. Both start with a 50\% belief that the other is High (moderate prior uncertainty). Figure 2 shows two scenarios where A’s belief about B evolves.

\begin{figure}[h!]
\centering
\includegraphics[width=0.5\textwidth]{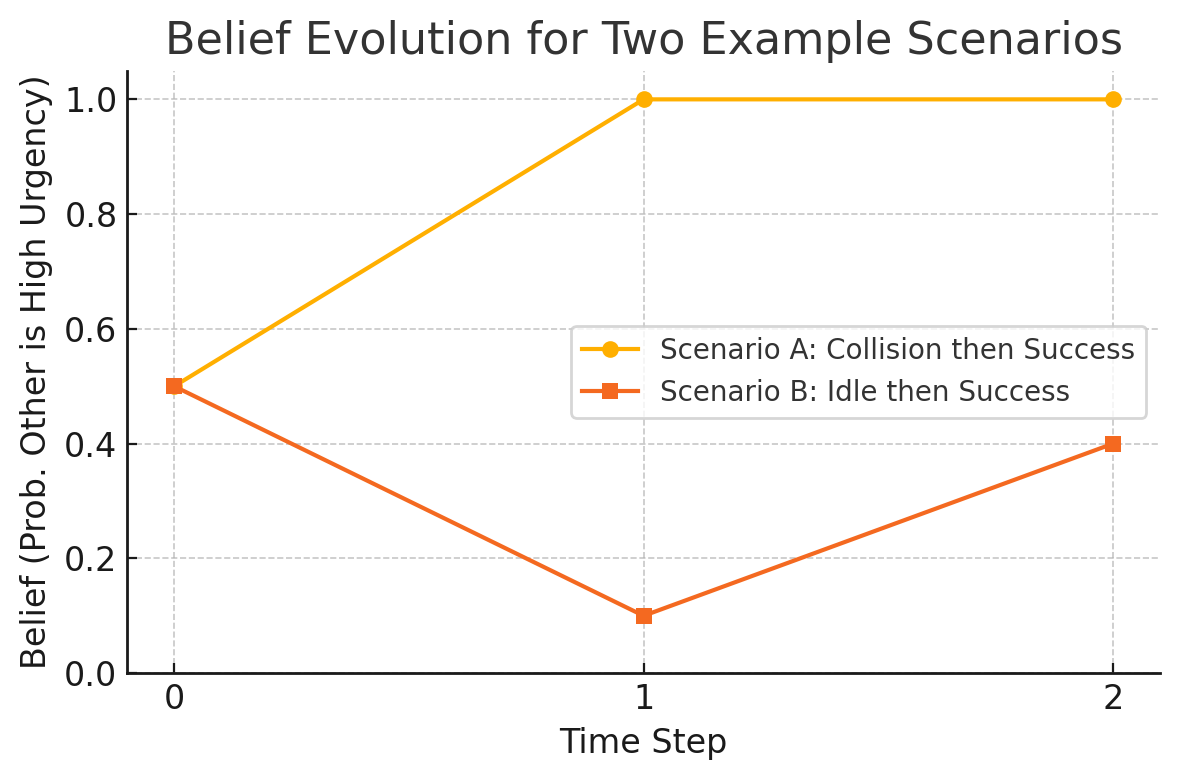}
\caption{Belief evolution for Agent A about Agent B's urgency in two scenarios. In Scenario A (yellow), a collision at $t=1$ (both dispatched) makes A's belief that B is High rise to 1.0 by $t=2$. In Scenario B (red), an idle slot at $t=1$ causes A’s belief in B being High to drop near 0, but a successful dispatch at $t=2$ slightly raises it.}
\end{figure}

In Scenario A, both A and B dispatch at $t=1$, causing a collision. A infers that B might actually be High urgency, as the likelihood of a Low urgency B dispatching simultaneously with High urgency A is low. Thus, $P(B=\text{High})$ increases to ~1.0 by $t=2$.

In Scenario B, neither agent dispatches at $t=1$ (idle slot). A suspects B is also Low urgency, reducing $P(B=\text{High})$. At $t=2$, A dispatches, and B does not. A now believes B is more likely Low, with only a slight increase in belief. By the end of Scenario B, A is confident that B is Low.

These scenarios show how common information (action history) refines agents' beliefs, improving coordination. After the collision in Scenario A, both agents know they are High urgency and randomize in the next slot (50\% dispatch). In Scenario B, A takes the lead after an idle start, preventing wasted slots. The symmetric strategy resolves uncertainty quickly, either through a collision or success, guiding agents to alternate or coordinate effectively. This aligns with the belief updates in Figure 2, reinforcing the policy’s learning-based design.

\textbf{Performance Comparison:} We now turn to the aggregate performance of the strategies. Figure 3 summarizes the expected total cost over $T=10$ slots under the optimal symmetric strategy versus a threshold heuristic and the always-dispatch policy, for the three load levels defined above (Light, Moderate, Heavy). These results are averaged over a large number of random trials for each condition, with the heuristic’s thresholds tuned (we chose $x=0.8, y=0.2, z=0.5$ as a reasonable performing set for the threshold policy in these experiments). The error bars (if shown) would be small due to many simulations, so we omit them for clarity.

\begin{figure}[h!]
\centering
\includegraphics[width=0.5\textwidth]{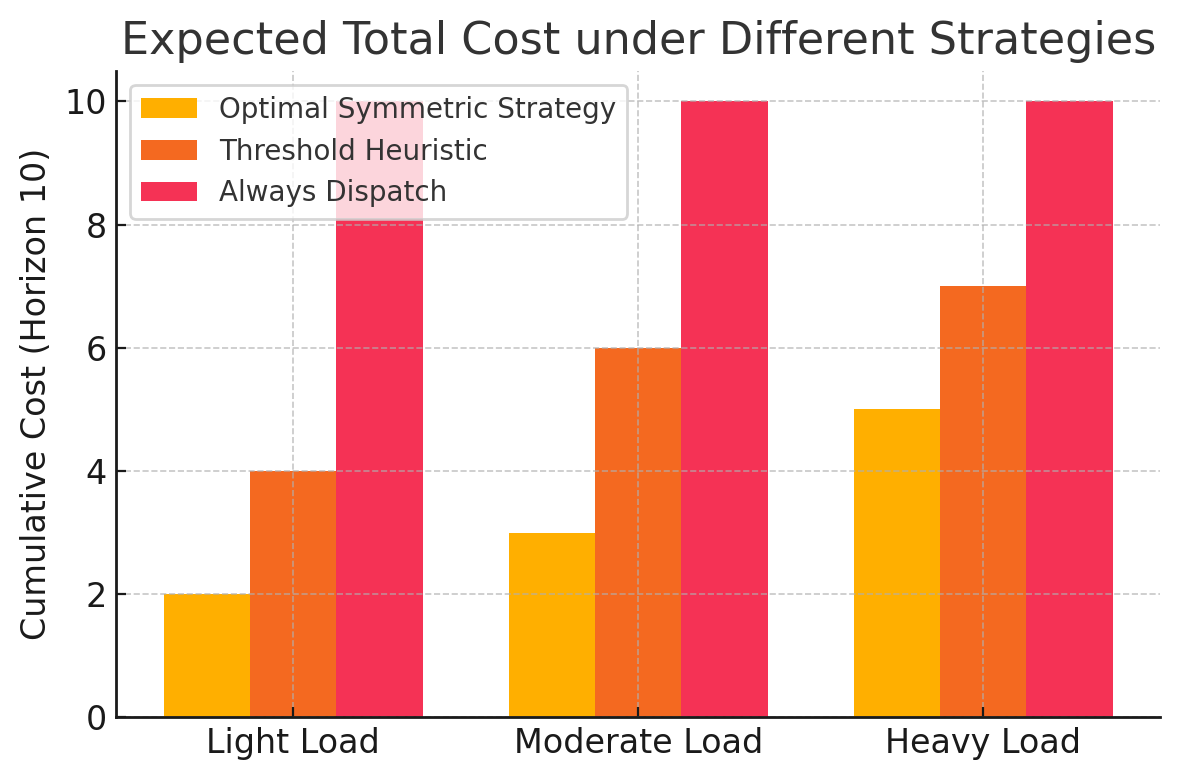}
\caption{Cumulative dispatch coordination cost (lower is better) under different strategies across load scenarios. Light Load: most slots have no urgent agents. Moderate Load: roughly one urgent agent at a time. Heavy Load: often multiple urgent agents competing. The optimal symmetric strategy achieves the lowest cost in all scenarios. The threshold-based heuristic improves over a naive “always dispatch” strategy, but still incurs higher costs (more collisions in heavy load, more idle in light load) compared to the proposed method.}
\end{figure}

Several trends are evident in Figure 3. First, the optimal symmetric strategy (yellow bars) consistently outperforms the alternatives. In the Heavy Load case, it cuts the total cost almost in half relative to the always-dispatch approach (from about 10 down to ~5 out of 10 possible penalty points). This means it successfully avoids many collisions even when both agents are frequently eager to send. In the Light Load case, it also achieves a low cost (around 2), indicating it prevents most idle slots despite agents rarely being urgent – effectively, the policy induces agents to take turns dispatching when needed so that nearly every slot is utilized by exactly one dispatch.

The always-dispatch policy (red bars) unsurprisingly fares worst in Light and Moderate loads: in Light load, always-dispatch causes unnecessary collisions in many slots where only one dispatch was needed; in Moderate, it still collides whenever two happen to be urgent together (which is fairly often). Notably, in Heavy Load, always-dispatch yields a cost of ~10, meaning every single slot resulted in a collision! This occurs because with both agents almost always urgent and always sending, they conflict every time (one gets through, one fails, but cost counts 1 for the conflict).

The threshold heuristic (orange bars) shows intermediate performance. It is better than always-dispatch in Heavy Load (cost ~7 vs 10) because the heuristic will sometimes hold one agent back if it believes the other is very likely to dispatch, reducing collisions. Similarly, in Light Load, it yields a cost of ~4 vs 10, because it encourages dispatch when likely idle and avoids having both go when not needed. However, the heuristic is not as fine-tuned as the optimal policy: for instance, in Moderate load its cost (~6) is significantly above the optimal (~3). The heuristic’s fixed threshold may cause it to err on the side of idleness or collision in borderline cases. In contrast, the optimal strategy essentially performs a soft thresholding – a delicate balance where dispatch probabilities smoothly adjust with beliefs, as illustrated earlier – leading to fewer mistakes.

In all cases, the symmetric DP strategy achieved the lowest expected cost (i.e., highest efficiency), a trend that held across all initial belief configurations we tested. An important qualitative observation is that the symmetric strategy maintains fairness in resource allocation. Because both agents use the same randomized policy, neither can consistently exploit the other; over many runs, each agent succeeds in dispatching a number of times proportional to its urgency level. We did not observe one agent getting an outsized share of dispatch slots in our simulations – a contrast to some naive policies where an aggressive agent might hog the channel. This fairness is inherent to the symmetry: if the agents were truly identical in their urgency distribution, their long-term outcomes are statistically indistinguishable. From a practical standpoint, this is a desirable property when coordinating independent organizations (e.g., different suppliers) since it encourages participation and trust in the protocol (no one feels disadvantaged).


\section{Conclusion and Future Work}
This paper presented a study on symmetric strategy optimization for dispatch coordination in a multi-agent supply chain, where distributed warehouses share a limited logistics resource. We formulated a model where all warehouses follow an identical decision rule based on private urgency and shared observations, using common information dynamic programming to convert the decentralized problem into a centralized one. This approach led to the provably optimal symmetric policy, which minimizes expected total cost by balancing delays and collision disruptions. 
Through simulations, we demonstrated that the optimal symmetric strategy outperforms baseline heuristics, achieving near-perfect slot utilization with minimal conflicts. In heavy demand scenarios, our method reduced collisions by 50

    
    



\textbf{Future Work:}
\begin{itemize}
    \item \textbf{Larger Networks:} Scaling to larger networks ($N \geq 10$) with approximation methods like mean-field control or neural networks for value function approximation.
    
    \item \textbf{Dynamic Urgencies:} Modeling evolving agent states as a partially observed Markov decision process, exploring reinforcement learning and robustness against deviations.
    
    \item \textbf{Incentive Analysis:} Investigating how the symmetric strategy forms a Bayesian Nash equilibrium, and designing mechanisms to realign incentives.
    
    \item \textbf{Prototype Deployment:} Testing the strategy with industry partners for practical feedback, particularly on common knowledge and probabilistic decisions.
    
    \item \textbf{Heterogeneous Resources:} Extending the strategy to allocate multiple resources (e.g., trucks or docks) while maintaining a symmetric threshold policy.
\end{itemize}
In conclusion, our approach offers a novel methodology for cooperative multi-agent strategies in supply chain coordination. By reframing the problem as a common-information control issue, we derived a solution that is efficient, fair, and highlights how distributed agents can self-organize for the common good. We hope this interdisciplinary approach inspires further research at the intersection of supply chain management, game theory, and control, fostering smarter and more autonomous logistics networks.

\section*{References}









\small

[1] Dominguez, R. \& Cannella, S.\ (2020) Insights on Multi-Agent Systems Applications for Supply Chain Management. {\it Sustainability} {\bf 12}(5):1935.

[2] GoRamp (2020) Congested Loading Docks -- Why Does It Happen? Online Article.

[3] Sudhakara, S. \& Nayyar, A.\ (2023) Optimal symmetric strategies in multi-agent systems with decentralized information. {\it arXiv preprint arXiv:2307.07150}.

[4] Sudhakara, S.\ (2023) Symmetric Strategies for Multi-Access IoT Network Optimization: A Common Information Approach. {\it arXiv:2310.15529}.

[5] Nayyar, A., Mahajan, A. \& Teneketzis, D.\ (2013) Decentralized stochastic control with partial history sharing: A common information approach. {\it IEEE Trans. Automatic Control} {\bf 58}(7).

[6] Wang, Z., et al.\ (2025) Research on collaborative scheduling strategies of multi-agent agricultural machinery groups. {\it Scientific Reports} {\bf 15}:9045.

[7]Sudhakara, S., Kartik, D., Jain, R. \& Nayyar, A.\ (2024) Optimal Communication and Control Strategies in a Cooperative Multiagent MDP Problem. {\it IEEE Trans. Automatic Control} {\bf 69}(10):6959--6966.

\end{document}